\ifx\pdfoutput\undefined
\documentstyle[prb,aps,amsmath,amssymb,multicol,graphicx]{revtex}
\else
\documentstyle[prb,aps,amsmath,amssymb,multicol,graphicx]{revtex}
\fi

\DeclareMathOperator{\EllipticK}{K}
\DeclareMathOperator{\EllipticE}{E}
\DeclareMathOperator{\JacobiSN}{sn}
\DeclareMathOperator{\JacobiDN}{dn}
\newcommand{\gtsd}{{\mathrm d}}
\newcommand{\gtsz}{z}
\newcommand{\Jacobim}{{\mathit m}}
\newcommand{\gtsu}{u}
\newcommand{\gtsSign}{\varepsilon}
\newcommand{\gtsConst}{{\mathsf c}}
\newcommand{\gtsCst}{c}
\newcommand{\gtsSurface}{{\mathcal S}}
\newcommand{\gtsHm}{{\mathcal H}}
\newcommand{\gtsJ}{{\mathit J}}
\newcommand{\gtsKc}{{\mathit k}_{c}}
\newcommand{\gtsJr}{\kappa}
\newcommand{\gtsMTheta}{\Theta}
\newcommand{\gtsMPhi}{\Phi}
\newcommand{\gtsq}{q}
\newcommand{\gtsQ}{Q}
\newcommand{\gtsm}{\widetilde{\Jacobim}}
\newcommand{\gtsSG}{\alpha}
\newcommand{\gtsK}{K}
\newcommand{\gtsE}{{\mathrm E}}
\newcommand{\gtsTE}{\underline{\mathrm E}}
\newcommand{\gtsEr}{{\mathcal E}}
\newcommand{\gtsTL}{\Lambda}
\newcommand{\gtsMC}{{\mathit H}}
\newcommand{\gtsLm}{\Jacobim}
\newcommand{\gtsLA}{A}
\newcommand{\gtsLB}{B}
\newcommand{\gtsLj}{j}
\newcommand{\gtsIL}{{\mathrm L}}
\newcommand{\gtsrd}{\lambda}

\draft
\title{Heisenberg Spins on a Cylinder Section
  \thanks{\textsc{Internat. J. Modern Phys. B} \textbf{14}, 2093 (2000) %
    [\texttt{cond-mat}/9909095]}}
\author{J.~Benoit$^\dag$, R.~Dandoloff$^\dag$ and A. Saxena$^\ddag$}
\address{%
  $^\dag$
  Laboratoire de Physique Th\'eorique et Mod\'elisation,
  Universit\'e de Cergy-Pontoise,\\
  95302 Cergy-Pontoise Cedex, France\\
  $^\ddag$%
  Theoretical Division, Los Alamos National Laboratory,\\
  Los Alamos, New Mexico 87545 USA
  }
\date{\texttt{cond-mat}/9909095}

\begin{document}
\maketitle
\begin{abstract}
Classical Heisenberg spins in the continuum limit (i.e. the
nonlinear $\sigma$-model) are studied on an elastic cylinder
section with homogeneous boundary conditions.  The latter may
serve as a physical realization of magnetically coated
microtubules and cylindrical membranes. The corresponding rigid
cylinder model exhibits topological soliton configurations with
geometrical frustration due to the finite length of the cylinder
section.  Assuming small and smooth deformations allows to find
shapes of the elastic support by relaxing the rigidity constraint:
an inhomogeneous Lam\'e equation arises.  Finally, this leads to
a novel geometric effect: a \textit{global shrinking} of the
cylinder section with swellings.
\end{abstract}

\pacs{75.10.Hk, 75.80.+q, 75.60.Ch, 11.10.Lm}

\begin{multicols}{2}
\section{Introduction}
Microtubules \cite{mic} and cylindrical membranes abound in nature.
They are important both in biological context and in industrial
applications.  The coupling of magnetism with the deformability of
these soft materials opens up new, exciting avenues of investigation.
We can envision these microtubules to be synthesized from either magnetic
organic materials or polymers \cite{magpol} or to enclose various
magnetorheological fluids \cite{fluids} in such a way that the magnetic
and elastic properties are coupled.  Specifically, the shape deformation
can be interpreted in terms of bending elasticity of the membrane.  To
explore magnetoelastic properties we will approximate the surfaces of
microtubules and cylindrical membranes as a continuum of classical
Heisenberg-coupled spins.

Our motivation is to amplify on the concept of geometrical frustration,
i.e. mismatch of different length scales in a system leading to
deformation of the geometry.  The length scales of interest are the
characteristic width of the spin (magnetic) soliton and the underlying
geometric length of the cylinder (i.e., radius). The main idea is that
the mismatch of length scales is reflected in the violation of the so
called self-duality equations and the increase in magnetic energy of
the system from that of the minimum energy in the corresponding homotopy
class.  Restoring self-duality or the minimum energy is achieved by
deforming the manifold in such a way that geometrical frustration is
relieved.

The continuum limit of the Heisenberg Hamiltonian for classical
ferromagnets or antiferromagnets for isotropic spin-spin coupling
is the nonlinear $\sigma$-model
\cite{BelavinPolyakov,Trimper,Chakravarty,Haldane,Fradkin}.
The total Hamiltonian for a deformable, magnetoelastically coupled
manifold is given by
$\gtsHm=\gtsHm_{magn}+\gtsHm_{el}+\gtsHm_{m-el}$,
where $\gtsHm_{magn}$, $\gtsHm_{el}$ and $\gtsHm_{m-el}$
represent the magnetic, elastic and magnetoelastic energy,
respectively.  In the present paper we will focus on the magnetic part
and the elastic part only since, for quasi-one-dimensional spin solutions
on the cylinder, $\gtsHm_{m-el}$ merely renormalizes $\gtsHm_{magn}$.
For the nonlinear $\sigma$-model, the magnetic energy
on a curved surface $\gtsSurface$, in curvilinear coordinates,
is given by
\cite{VSDTS,TSGF}
\begin{equation}
\label{Hm/mg}
\gtsHm_{magn}=
\gtsJ\!{\iint_\gtsSurface}\!\sqrt{g}\,{\gtsd\Omega}\:
g^{ij}h_{\alpha\beta}\partial_{i}n^{\alpha}\partial_{j}n^{\beta},
\end{equation}
where $\gtsJ$ denotes the coupling energy between neighbouring spins.
The order parameter $\widehat{{\mathbf n}}$ is the local magnetization
unit vector specified by a point on the sphere $S^2$. The metric
tensors $(g_{ij})$ and $(h_{\alpha\beta})$ describe respectively
the support surface $\gtsSurface$ and the order parameter manifold:
as customary, ${\gtsd\Omega}$ represents the surface area and
$g$ the determinant $\det(g_{ij})$.

\section{Rigid cylinder section:\\ non-self dual solitons}
First, let us consider the nonlinear $\sigma$-model
on a rigid cylinder section.

\subsection{Formulation}
For our purposes a suitable representation,
in cylindrical coordinates $(\rho,\varphi,z)$,
is given by
\begin{equation}
\rho=r,
\end{equation}
where the constant parameter $r$ is real and positive,
while the angle $\varphi$ varies from $-\pi$ to $\pi$.
One can easily check that the metric is given by
\begin{equation}
g=r^2\ \gtsd\varphi\!\otimes\!\gtsd\varphi+%
  \gtsd\gtsz\!\otimes\!\gtsd\gtsz,
\end{equation}
therefore $g^{\varphi\gtsz}=g^{\gtsz\varphi}=0$ and we have
\begin{equation}
\label{T/g/nested}
g^{\varphi\varphi}\!\sqrt{g}=\frac{1}{r},
\quad g^{\gtsz\gtsz}\!\sqrt{g}={r}.
\end{equation}
From now on, we restrict ourselves to a section of the cylinder:
the $\gtsz$ variable will vary from $-\Delta\gtsz$ to $\Delta\gtsz$
where \mbox{$0\leqslant\Delta\gtsz\leqslant\infty$}.

As usual,
the local magnetization $\widehat{{\mathbf n}}$ is described
by its spherical coordinates $(\gtsMTheta,\gtsMPhi)$,
then the metric on the Heisenberg sphere is given by
\begin{equation}
h=\gtsd\gtsMTheta\!\otimes\!\gtsd\gtsMTheta
+\sin^2 \gtsMTheta\ \gtsd\gtsMPhi\!\otimes\!\gtsd\gtsMPhi.
\end{equation}

Assuming homogeneous boundary conditions
at both ends \mbox{($\gtsMTheta=0\left[\pi\right]$}
as \mbox{$\gtsz\to\pm\Delta\gtsz$)}
allows to map each boundary of the section to a point:
thus we compactify our cylinder section into the sphere $S^2$.
Consequently, the mapping of our support to the order parameter
manifold is classified by the homotopy group $\Pi_2\left(S^2\right)$
which is isomorphic to $\mathbb{Z}$:
spin configurations may be classified according to
their homotopy class \cite{BelavinPolyakov,Bogomolnyi}.

\subsection{Derivation of non-self dual solitons}
Henceforth, without loss of generality,
only cylindrical symmetric configurations
\mbox{($\partial_{\varphi}\gtsMTheta=\partial_{\gtsz}\gtsMPhi=0$)}
will be considered.
Thus the magnetic Hamiltonian (\ref{Hm/mg}) becomes
\begin{equation}
\label{Hm/mg/xi}
\gtsHm_{magn}=
\gtsJ\!
\int\limits_{-\Delta\gtsz}^{+\Delta\gtsz}\!\!\gtsd\gtsz\!\!
\int\limits_{-\pi}^{+\pi}\!\!\gtsd\varphi\,
\left[
{r}{\gtsMTheta_{\gtsz}^2}
+\frac{\sin^2 \gtsMTheta}{r} \gtsMPhi_{\varphi}^2
\right],
\end{equation}
where a subscript stands for differentiation.
Rescaling the $z$ variable
in equation (\ref{Hm/mg/xi}) gives:
\begin{equation}
\label{Hm/mg/zt}
\gtsHm_{magn}=
\gtsJ\!
\int\limits_{-\Delta\zeta}^{+\Delta\zeta}\!\!\gtsd\zeta\!\!
\int\limits_{-\pi}^{+\pi}\!\!\gtsd\varphi\,
\left[
\gtsMTheta_{\zeta}^2 +\sin^2 \gtsMTheta\ \gtsMPhi_{\varphi}^2
\right],
\end{equation}
where $\zeta\equiv\gtsz/r$ and
$\Delta\zeta\equiv \Delta\gtsz/r$.
The Euler-Lagrange equations
corresponding to (\ref{Hm/mg/zt}) are:
\begin{subequations}
\label{EL/Hm/mg/zt}
\begin{eqnarray}
\gtsMPhi_{\varphi\varphi}&=&0,\label{EL/Hm/mg/zt/MPhi}\\
\gtsMTheta_{\zeta\zeta}&=&
\gtsMPhi_{\varphi}^2 \sin \gtsMTheta \cos \gtsMTheta.
\label{EL/Hm/mg/zt/MTheta}
\end{eqnarray}
\end{subequations}
\begin{subequations}
\label{EL/Hm/mg/zt/nested}
From (\ref{EL/Hm/mg/zt/MPhi}), it follows that
\begin{equation}
\label{EL/Hm/mg/zt/MPhi/nested}
\gtsMPhi_{\varphi}=q_{\varphi}\qquad q_{\varphi}\in\mathbb{Z}.
\end{equation}
Substituting this into (\ref{EL/Hm/mg/zt/MTheta}) and
rescaling again the rotating angle, we get the
sine-Gordon (\textsc{sg}) equation
\begin{equation}
\label{SG/MTheta}
\gtsMTheta_{\varrho\varrho}=\sin \gtsMTheta \cos \gtsMTheta,
\end{equation}
where $\varrho\equiv \gtsq_{\varphi}\zeta$.
\end{subequations}

Equation (\ref{SG/MTheta}) may be integrated once to yield
\begin{equation}
\label{SG/res/int1}
\gtsMTheta_{\varrho}^2=\sin^2 \gtsMTheta +\gtsm
\qquad \gtsm\in\left[0,+\infty\right).
\end{equation}
Performing the change of variable
$\sin\gtsMTheta=\JacobiDN\!\left(\gtsu\mid 1+\gtsm\right)$
where $\JacobiDN$ is a Jacobi elliptic function \cite{Abramowitz},
the differential equation becomes $\gtsu_{\varrho}^2=1$.
Let us denote by $\gtsSG\!\left(\cdot\mid\gtsm\right)$
the increasing solution of (\ref{SG/MTheta})
specified by the \emph{parameter} $\gtsm$ and subject to
the boundary condition $\gtsSG\!\left(0\mid\gtsm\right)=\frac{\pi}{2}$.
Readily, we get:
\begin{equation}
\label{SG/res/Jacobi/sin}
\sin\gtsSG\!\left(\varrho\mid\gtsm\right)=
\JacobiDN\!\left(\varrho\mid 1+\gtsm\right).
\end{equation}
Thus, the general \textsc{sg} solution, in natural coordinates, is
\begin{equation}
\label{SG/res/gsol}
\theta\!\left(\varrho\right)=
\gtsSign\gtsSG\!\left(\varrho\mid\gtsm\right)+\gtsConst
\qquad\gtsSign=\pm{1}.
\end{equation}
Further the function $\gtsSG\!\left(\cdot\mid\gtsm\right)$ satisfies
\begin{equation}
\label{SG/rel/qp}
\gtsSG\!\left(\varrho+2\gtsq\gtsK_{\gtsSG}\mid\gtsm\right)=
\gtsSG\!\left(\varrho\mid\gtsm\right)+\gtsq\pi
\qquad\gtsq\in\mathbb{Z},
\end{equation}
where the \emph{quasi quarter-period} $\gtsK_{\gtsSG}$
is related to the complete elliptic integral of
the first kind $\EllipticK$ by
\cite{Abramowitz}
\begin{subequations}
\label{SG/qp/ei}
\begin{eqnarray}
\label{SG/qp/ei/crude}
\gtsK_{\gtsSG}\!\left(\gtsm\right)
&=&\EllipticK\!\left(1\!+\!\gtsm\right),\\
\label{SG/qp/ei/nested}
&=&\frac{1}{\sqrt{1\!+\!\gtsm}}\:
  \EllipticK\!\!\left(\frac{1}{1\!+\!\gtsm}\right).
\end{eqnarray}
\end{subequations}
Clearly,
the parameter $\gtsm$ tunes the quasi quarter-period
$\gtsK_{\gtsSG}$.

Using the solutions of equation (\ref{SG/MTheta}),
the \mbox{$\gtsq_{\zeta}\pi$}-soliton configuration consistent with the
boundary conditions can be obtained easily.
Up to an irrelevant additive multiple of $\pi$, we have
\begin{subequations}
\label{SConf}
\begin{equation}
\label{SConf/MTheta}
\gtsMTheta\!\left(\zeta\right)=
\gtsSign\gtsSG\!\left(\gtsq_{\varphi}\zeta\mid\gtsm\right)
-\delta_{\text{even},\gtsq_{\zeta}}{\textstyle{\frac{\pi}{2}}},
\end{equation}
where the parameter $\gtsm$ is given by
\begin{equation}
\label{SConf/m}
\gtsm=\gtsK_{\gtsSG}^{-1}\!\!\left(
\frac{\gtsq_{\varphi}}{\gtsq_{\zeta}}\Delta\zeta
\right).
\end{equation}
\end{subequations}
If we consider an infinite cylinder,
the solution (\ref{SConf}) represents a soliton lattice
with a period $2\gtsK_{\gtsSG}$.
When $\gtsm$ tends to $0^{+}$,
this period tends to $\infty$
and we recover the self-dual equations\cite{TSGF}.

\subsection{Geometric frustration}
The magnetic energy $\gtsE_{magn}$ of the above configuration (\ref{SConf})
may be compared with the corresponding topological minimum energy $\gtsTE$
which does not depend on the geometry of the support manifold.
Performing the Bogomol'nyi's decomposition\cite{Bogomolnyi} yields
\begin{equation}
\label{Bgml/E}
\gtsTE=8\pi\gtsJ\vert\gtsQ\vert,
\end{equation}
where the topological charge (i.e. the winding number)
$\gtsQ$ equals to $\gtsq_{\zeta}\gtsq_{\varphi}$.
Let $\gtsEr_{magn}$ denote the ratio $\gtsE_{magn}/\gtsTE$.
A straightforward calculation shows that $\gtsEr_{magn}$ depends
on the parameter $\gtsm$ only; we have
\begin{subequations}
\label{Bgml/Er}
\begin{eqnarray}
\gtsEr_{magn}
\label{Bgml/Er/crude}
&=&\EllipticE\!\left(1+\gtsm\right)+
  {\textstyle\frac{1}{2}}\gtsm\:
  \EllipticK\!\left(1+\gtsm\right),\\
\label{Bgml/Er/nested}
&=&\sqrt{1+\gtsm}\left[
  \EllipticE\!\!\left(\frac{1}{1\!+\!\gtsm}\right)\!-\,
  {\textstyle\!\frac{1}{2}}\frac{\gtsm}{1\!+\!\gtsm}\:
  \EllipticK\!\!\left(\frac{1}{1\!+\!\gtsm}\right)
\right],
\end{eqnarray}
\end{subequations}
which increases strictly from $1$ to $\infty$
with respect to $\gtsm$;
$\EllipticE$ is the complete elliptic integral of
the second kind\cite{Abramowitz}.
Therefore, according to (\ref{SConf/m}) $\gtsEr_{magn}$
is increasing function of the radius $r$
and decreasing function of $\Delta\gtsz$.

\section{Elastic cylinder section: frustration release}
Next, let us relax the rigidity constraint and consider
the nontrivial spin configuration on an elastic cylinder section.
Accordingly, the soliton will try to minimize its
magnetic energy $\gtsE_{magn}$ to the minimum energy $\gtsTE$
by deforming the elastic support\cite{VSDTS,TSGF,CIGFMS}.

\subsection{Deformable metric}
Since the radius $r$ appears as the relevant
geometric parameter and $\gtsz$ variable as the relevant
curvilinear coordinate, we relax $r$ with respect to $\gtsz$ and
write
\begin{equation}
\label{aT/ea}
r\!\left(\gtsz\right)=r_{0}\:\left[1+\gtsTL\!\left(\gtsz\right)\right],
\end{equation}
where $r_{0}$ represents the \emph{spontaneous radius}
and the function $\gtsTL$ describes local deformations.
The metric on the deformable manifold in this case
remains orthogonal, and (\ref{T/g/nested}) reads
\begin{equation}
\label{aT/g/nested}
g^{\varphi\varphi}\!\sqrt{g}\!=
  \!\!\frac{\sqrt{1+r_{0}^2\gtsTL_{\gtsz}^2}}{r},
\quad
g^{\gtsz\gtsz}\!\sqrt{g}\!=
  \!\!\frac{r}{\sqrt{1+r_{0}^2\gtsTL_{\gtsz}^2}}.
\end{equation}

\subsection{Derivation of small and smooth deformations}
As the problem is quasi-one-dimensional,
the magnetoelastic energy $\gtsHm_{m-el}$ merely renormalizes
the spin coupling energy $\gtsJ$ in the magnetic energy
$\gtsHm_{magn}$\cite{CHC}.
Therefore, we add to the nonlinear
$\sigma$-model Hamiltonian (\ref{Hm/mg})
only the elastic energy which is essentially
stored in the bending of the deformable support
\cite{Canham,HelfrichZN1,MAPeterson,SMMS}:
\begin{equation}
\label{Hm/el}
\gtsHm_{el}={\textstyle{\frac{1}{2}}}\gtsKc\!
{\iint_\gtsSurface}\!\sqrt{g}\,{\gtsd\Omega}\:
\left(\gtsMC-\gtsMC_{0}\right)^2.
\end{equation}
Here the constant $\gtsKc$ denotes the \emph{bending rigidity} of
the (cylinder) material,
$\gtsMC$ represents the mean curvature \cite{Struik},
and the \emph{spontaneous mean curvature} $\gtsMC_{0}$ tends
to bias the mean curvature for recovering the spontaneous shape.
Assuming small and smooth deformations and expanding to second order
in $\gtsTL$, $\gtsTL_{\gtsz}$ and $\gtsTL_{\gtsz\gtsz}$
lead to
\begin{equation}
\label{aT/Hm/el}
\gtsHm_{el}=
{\textstyle{\frac{1}{8}}}\pi\gtsKc\!
\int\limits_{-\Delta\zeta}^{+\Delta\zeta}\!\!\!\gtsd\zeta\,\gtsTL^2.
\end{equation}

Before deriving the Euler-Lagrange equation for the total
Hamiltonian $\gtsHm=\gtsHm_{magn}+\gtsHm_{el}$,
we calculate the magnetic energy associated with
the nontrivial spin configuration (\ref{SConf}).
Expanding to second order in $\gtsTL$ and $\gtsTL_{\gtsz}$ the
relations (\ref{aT/g/nested}) enable to rewrite (\ref{Hm/mg/zt})
as follows
\begin{eqnarray}
\label{aT/Hm/mg}
\gtsHm_{magn}&&=
\gtsJ\!
\int\limits_{-\Delta\zeta}^{+\Delta\zeta}\!\!\gtsd\zeta\!\!
\int\limits_{-\pi}^{+\pi}\!\!\gtsd\varphi\,
\left[\vphantom{\frac{1}{2}}
\left[\gtsMTheta_{\zeta}^2 +\sin^2 \gtsMTheta\ \gtsMPhi_{\varphi}^2
\right]\right.\nonumber\\
-&&\left(-\gtsTL+{\textstyle\frac{1}{2}}\gtsTL_{\zeta}^2\right)
\left[\gtsMTheta_{\zeta}^2 -\sin^2 \gtsMTheta\ \gtsMPhi_{\varphi}^2
\right]\nonumber\\
&&\left.\hphantom{\gtsJ\!\int\!\!\gtsd\zeta\int\!\!\gtsd}
+{\gtsTL^2\:\sin^2 \gtsMTheta\ \gtsMPhi_{\varphi}^2}
\vphantom{\frac{1}{2}}
\right].
\end{eqnarray}
On the other hand,
according to (\ref{SG/res/int1}) and (\ref{SG/res/Jacobi/sin}),
the spin configuration (\ref{SConf}) verifies
\begin{eqnarray*}
\gtsMTheta_{\zeta}^2&-&\sin^2\gtsMTheta\ \gtsMPhi_{\varphi}^2=
\gtsm\gtsq_{\varphi}^2,\\
\gtsMTheta_{\zeta}^2&=&
\left(1+\gtsm\right)\gtsq_{\varphi}^2
\left[1-\JacobiSN^2\!\!\left(\gtsq_{\varphi}\zeta\mid 1+\gtsm\right)
\right].
\end{eqnarray*}
Inserting these relations into (\ref{aT/Hm/mg}) and simplifying,
we obtain
\begin{eqnarray}
\label{SConf/Hm/mg/2nd}
\gtsHm_{magn}&&=
2\pi\gtsJ\gtsq_{\varphi}^2\!
\int\limits_{-\Delta\zeta}^{+\Delta\zeta}\!\!\gtsd\zeta\,
\left[\vphantom{\frac{1}{2}}
\left(2+\gtsm\right)-\gtsm
\left[-\gtsTL+{\textstyle\frac{1}{2}}\gtsTL_{\zeta}^2
\right]\right.\nonumber\\
-&&\left.
\left(1+\gtsm\right)
\left[2+\gtsTL^2\right]
\JacobiSN^2\!\!\left(\gtsq_{\varphi}\zeta\mid 1+\gtsm\right)
+\gtsTL^2
\vphantom{\frac{1}{2}}
\right].
\end{eqnarray}
The Euler-Lagrange equation for the problem takes the following form
\begin{subequations}
\label{EL/TL}
\begin{equation}
\label{EL/TL/eqn}
\gtsTL_{\varrho\varrho}+
\left[\left(1+\gtsLm\right)\!\gtsLA
-\gtsLm\gtsLB\;\JacobiSN^2\!\!\left(\varrho\mid\gtsLm\right)
\right]\gtsTL
=\sqrt{\gtsLm}\ \gtsLj,
\end{equation}
where we have set
\begin{eqnarray}
\gtsLm&=&1+\gtsm,\\
\gtsLA&=&\frac{2}{\gtsm\gtsq_{\varphi}^2}
\left[1+\frac{1}{16}\:\frac{\gtsCst_{2}}{\gtsJr\gtsq_{\varphi}^2}
\right],\\
\gtsLB&=&\frac{2}{\gtsm\gtsq_{\varphi}^2},\\
\gtsLj&=&\frac{-1}{\gtsq_{\varphi}^2\sqrt{1+\gtsm}}.
\end{eqnarray}
\end{subequations}
Here we have introduced the \emph{relative coupling energy}
$\gtsJr\equiv\gtsJ/\gtsKc$ and the variable
$\varrho\equiv\gtsq_{\varphi}\zeta$ (see (\ref{SG/MTheta})).
The linear inhomogeneous second-order differential equation
(\ref{EL/TL/eqn}) is related to the well-known homogeneous
(Jacobian) Lam\'e's equation which occurs in several physical
contexts \cite{HST,Arscott}.
We have found no direct treatment of
(\ref{EL/TL/eqn}) in the literature.
However, an approach based on the derivation of
the Lam\'e functions \cite{HST,Arscott}
allows to find a particular solution denoted by
$\gtsIL\!\left(\cdot\mid\gtsLm;\gtsLA,\gtsLB,\gtsLj\right)$.
Therefore, if small and smooth deformations are assumed, a
suitable deformation function $\gtsTL$ is given by
\begin{equation}
\label{aT/TL}
\gtsTL\!\left(\zeta\right)=
\gtsIL\!\left(
\gtsq_{\varphi}\ \zeta\mid 1+\gtsm;\gtsLA,\gtsLB,\gtsLj
\right).
\end{equation}

\subsection{Frustration release}
The geometric frustration becomes evident in considering
the Euler-Lagrange equation (\ref{EL/TL/eqn}):
$\gtsTL=0$ (i.e. the rigid cylinder section) is not a solution.
Now let us turn our attention to the deformation mechanism.
As we have seen, the magnetic energy $\gtsE_{magn}$
of the configuration (\ref{SConf}) is
an increasing function of the radius $r$:
therefore for sufficiently small and smooth deformations,
the soliton tries to decrease the radius $r$.
In other words, the soliton tends to collapse
the cylinder section.
On the other hand,
the elastic Hamiltonian (\ref{Hm/el}) tries to maintain
the spontaneous shape of the deformable support.
Consequently, the competition between the magnetic energy
and the elastic energy induces a decrease of the radius $r$.
Since according to (\ref{Hm/mg/zt}) the soliton energy
is essentially localized in the spread zone,
the deformation is less important in the spin flip region.

\begin{figure}[t]
  \begin{center}
    \includegraphics[width=\linewidth]{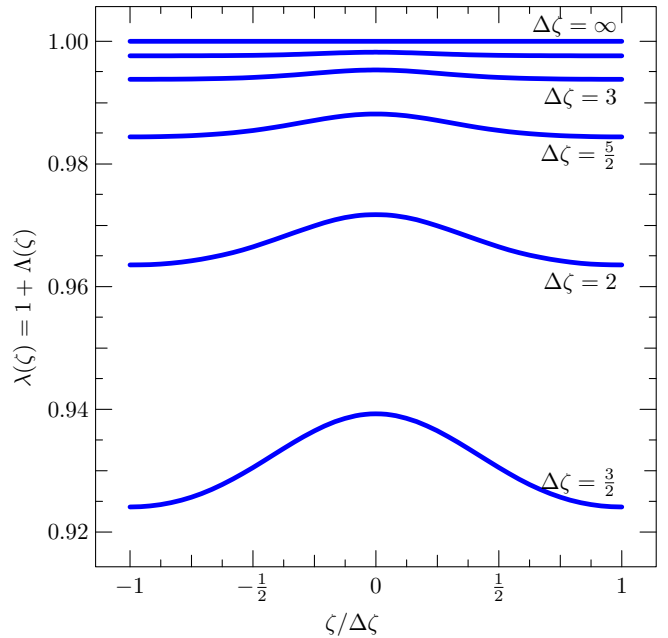}
  \end{center}
\caption[relative dilatation]{%
  The relative dilatation $\gtsrd$ as defined
  in (\ref{FR/rd/def}) corresponding to the deformation
  function (\ref{aT/TL}) associated with cylinder sections
  in presence of a $2\pi$-soliton
  versus $\zeta/\Delta\zeta$ for different values of
  $\Delta\zeta$
  and with the relative coupling energy $\kappa$
  fixed to $1/16$.
  }
\label{fig/RDPC/k}
\end{figure}

To understand the geometric meaning of the frustration release,
let us define the \emph{relative dilatation} $\gtsrd$ by
\begin{equation}
\label{FR/rd/def}
\gtsrd\equiv\frac{r}{r_{0}}=1+\gtsTL.
\end{equation}
As shown in Fig.~\ref{fig/RDPC/k},
the decrease of the radius leads to
a global shrinking whereas a local swelling arises
where the spins twist.

\section{Conclusion}
In conclusion, we showed that cylindrical magnetic surfaces, if
approximated by classical Heisenberg spins, exhibit novel
magnetoelastic deformation effects in the presence of spin solitons
due to a mismatch of length scales.  In particular, we showed that
the Euler-Lagrange equation for the nonlinear $\sigma$-model on a
cylinder section with homogeneous boundary conditions is the
sine-Gordon equation.  As announced, the frustration release leads
to a novel geometric effect: the cylinder section is
\textit{globally shrunk} and a swelling appears in the region of
the soliton.

\section*{Acknowledgement}
This work was supported in part by the US Department of Energy.


\end{multicols}
\end{document}